\documentclass[fleqn,usenatbib]{mnras}

\usepackage{newtxtext,newtxmath}

\usepackage[T1]{fontenc}
\usepackage{ae,aecompl}

\usepackage{graphicx}	
\usepackage{amsmath}	
\usepackage{amssymb}	
\usepackage{color}

\newcommand{\integral}{\textit{INTEGRAL}}

\newcommand{\nustar}{\textit{NuSTAR}}

\newcommand{\swift}{{\it Swift}}

\newcommand{\xmm}{{\it XMM-Newton}}

\newcommand{\kms}{km~s$^{-1}$}
\newcommand{\pcm}{cm$^{-2}$}
\newcommand{\src}{NGC~1566}

\title[X-ray spectra of NGC 1566]{X-ray spectra reveal the reawakening of the repeat changing-look AGN NGC 1566}

\author[M. L. Parker et al.]{M. L. Parker,$^{1}$\thanks{E-mail: mparker@sciops.esa.int}
N. Schartel,$^{1}$
D. Grupe,$^{2}$
S. Komossa,$^{3}$
F. Harrison,$^{4}$\newauthor
W. Kollatschny,$^{5}$ 
R. Mikula,$^{2}$
M. Santos-Lle\'{o}$^{1}$
and L. Tom\'{a}s$^{1}$
\\
$^{1}$European Space Agency (ESA), European Space Astronomy Centre (ESAC), E-28691 Villanueva de la Ca\~{n}ada, Madrid, Spain\\
$^{2}$Department of Earth and Space Sciences, Morehead State University, Morehead, KY 40514, USA\\
$^{3}$Max-Planck-Institut f\"{u}r Radioastronomie, Auf dem H\"{u}gel 69, 53111 Bonn, Germany\\
$^{4}$Cahill Center for Astrophysics, 1216 E. California Blvd, California Institute of Technology, Pasadena, CA 91125, USA\\
$^{5}$Institut f\"{u}r Astrophysik, Universit\"{a}t G\"{o}ttingen, Friedrich-Hund Platz 1, D-37077 G\"{o}ttingen, Germany\\
}

\date{Accepted XXX. Received YYY; in original form ZZZ}

\pubyear{2015}

\begin{document}
\label{firstpage}
\pagerange{\pageref{firstpage}--\pageref{lastpage}}
\maketitle

\begin{abstract}
We present simultaneous \xmm\ and \nustar\ observations of the repeat changing-look AGN \src , which dramatically increased in brightness in the IR to X-ray bands in 2018. The broad-band X-ray spectrum was taken at the peak of the outburst and is typical of Seyfert~1 AGN. The spectrum shows a soft excess, Compton hump, warm absorption and reflection, ruling out tidal disruption as the cause of the outburst and demonstrating that a `standard' accretion disk can develop very rapidly. The high resolution grating spectrum reveals that the outburst has launched a $\sim500$~\kms\ outflow, and shows photoionised emission lines from rest-frame gas. We discuss possible mechanisms for the outburst, and conclude that it is most likely caused by a disk instability.
\end{abstract}




\section{Introduction}

Active Galactic Nuclei (AGN) exhibit large changes in flux across many bands \citep[e.g. review by][]{Uttley14_multiwavelength}. In X-rays in particular, many AGN are known to change in luminosity by orders of magnitude on timescales from months to a few hours \citep[e.g.][]{Boller97, Komossa17_extremeAGN}. We have an ongoing observing program studying AGN in anomalous flux states, which has been successful in revealing unusual reflection-dominated states \citep{Schartel07,Grupe12}, strong absorption events \citep{Grupe13,Parker14_mrk1048}, and bright outbursts \citep{Parker16_he1136}. 
Of particular interest among AGN with extreme variability are the changing-look AGN, which change their Seyfert classification \citep[e.g.][]{Penston84} and are sometimes associated with a switch from Compton-thick to Compton-thin absorption in the X-ray band \citep{Guainazzi02, Matt03}. 

\src\ is a local ($z=0.005$) face-on Seyfert galaxy, which was observed to increase dramatically in flux in 2018.
This activity in \src\ was detected serendipitously by \integral\ \citep{NGC1566_atel1} and followed up with the \emph{Neil Gehrels Swift Observatory} (\swift), which found it to be a factor of $\sim15$ brighter than archival observations in X-rays \citep[e.g.][]{Kawamuro13} and nearly 3 magnitudes brighter in the UVW2 filter \citep{NGC1566_atel3,NGC1566_atel7}. The \emph{ASAS-SN} optical and \emph{NEOWISE} infra-red lightcurves show that the source has been brightening since September 2017 \citep{NGC1566_atel4,NGC1566_atel5}, and a SAAO optical spectrum showed \citep{NGC1566_atel6} much stronger broad emission lines, consistent with a change in Seyfert type to 1.2 from its typical quiescent Sy~1.9--1.8 type \citep{Oknyansky18}. Interestingly, these outbursts are recurrent: \citet{Alloin86} identify four separate periods of activity between 1970 and 1985, each lasting for $\sim1300$ days and with associated increases in broad-line strength causing the Seyfert type to move between Sy2 and Sy1. Another outburst in 2010 is visible in the \swift\ Burst Alert Telescope (BAT) 105-month lightcurve\footnote{\url{https://swift.gsfc.nasa.gov/results/bs105mon/216}}.

In this letter, we present broadband X-ray spectroscopy of the peak of the 2018 outburst of \src\ with \xmm\ and \nustar .

\section{Observations and data reduction}
 
\subsection{XMM-Newton}
Based on the detection of enhanced X-ray activity by \integral , we triggered a joint \xmm\ \nustar\ target of opportunity (ToO) observation (\xmm\ proposal ID 080084, PI Schartel). The observation length was 94~ks, taken on June 26 2018 (obs. ID 0800840201).
We reduce the \xmm\ data with the science analysis software (SAS) version 16.1.0. 
We reduce the Reflection Grating Spectrometer (RGS) data using the SAS task \emph{rgsproc}. We filter the data for background flaring using a threshold of 0.2 counts~s$^{-1}$. We combine the 1st and 2nd order spectra and the spectra from the two detectors into a single spectrum using \emph{rgsfluxcombine}, and convert this to \textsc{spex} format using \emph{rgsfmt}. We fit the RGS data from 8--35\AA. 
Due to the source's low redshift, the EPIC count rates are very high (40~s$^{-1}$ in the pn), and the data is piled-up. To mitigate this, we use only the least affected EPIC-pn data, and use an annular source region. There are several ultra-luminous X-ray sources (ULXs) in NGC~1566 \citep{Liu05}, but these are outside the window or obscured by the PSF of the AGN, which dominates the total count rate.
We process the pn data using the \emph{epproc} tool, and filter for background flares, leaving a clean exposure time of 65~ks. We extract source photons from a 30$^{\prime\prime}$ radius annulus centered on the source, with an inner radius of 8$^{\prime\prime}$, and a 40$^{\prime\prime}$ background region, extracted from the furthest corner of the detector. We bin the EPIC spectrum to oversample the data by a factor of 3, and to a minimum signal to noise ratio of 6. We fit the pn data from 0.5--10~keV, excluding the 2--2.5~keV band where there is a calibration feature \citep[see appendix of][]{Marinucci14}.

We also reduce the data from the observation of \src\ taken in 2015 when the source was faint, using the same procedure. A full analysis of these data will be presented in Tom\'{a}s et al. (in prep.).

\subsection{NuSTAR}
A NuSTAR observation of 80~ks was taken simultaneously with the \xmm\ exposure (obs. ID 80301601002). 
We reduced the \nustar\ data using the \nustar\ Data Analysis Software (NuSTARDAS) version 1.6.0. We extract source counts from a 60$^{\prime\prime}$ circular extraction region, and background counts from a 90$^{\prime\prime}$ circular extraction region on the same chip. We bin the spectra to a signal to noise ratio of 6, and to oversample the instrumental resolution by a factor of 3. We fit the FPMA and FPMB spectra separately, but group them in \textsc{xspec} for plotting purposes.

\subsection{Swift}
After the flare was detected \swift\ followup observations were immediately requested. The X-ray Telescope \citep[XRT,][]{Burrows05} observations were mostly performed in Windowed Timing (WT) mode \citep[][]{hill05}, however some initial observations were performed in photon counting (pc) mode. The XRT data were reduced 
using the task {\it xrtpipeline}.
Background and source events were extracted with \textsc{xselect}. For the WT data we used a 40 by 3 pixel box, rotated to match the spectrum orientation. Because the first observations after the flare were in pc mode they were strongly affected by pileup, so we excluded the inner part of the PSF using an annular extraction region with inner and outer radii of 16.5" and 94.3". 
We used the latest (2013) response files, and created auxiliary response files (ARFs) with the FTOOL {\it xrtmkarf}. We binned the spectra with 20 counts per bin. 
Typical exposure times per spectrum were of the order of 1ks. 

The UVOT data of each observation were coadded in each filter. 
We extracted source counts from an extraction region with a radius of 3". The loss in the PSF was corrected with the command {\it uvotsource}. 
Count to flux density and magnitude conversion was performed based on the most recent calibration files \citep{Poole08, Breeveld10}. The UVOT data were corrected for Galactic reddening \citep[$E_{\rm B-V}=0.025$;][]{Schlegel98}.

\begin{figure*}
\centering
\includegraphics[height=5.8cm]{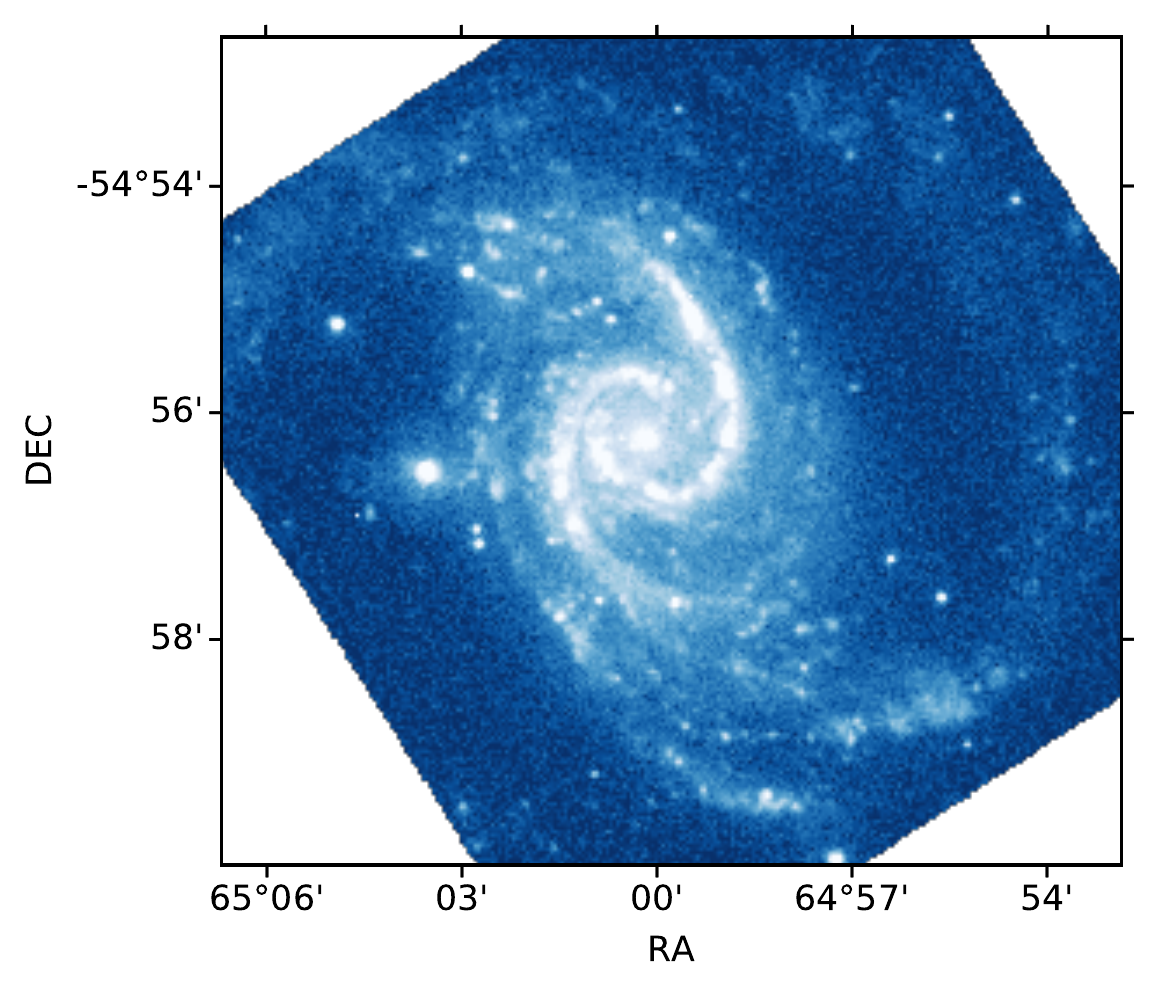}
\includegraphics[height=5.8cm]{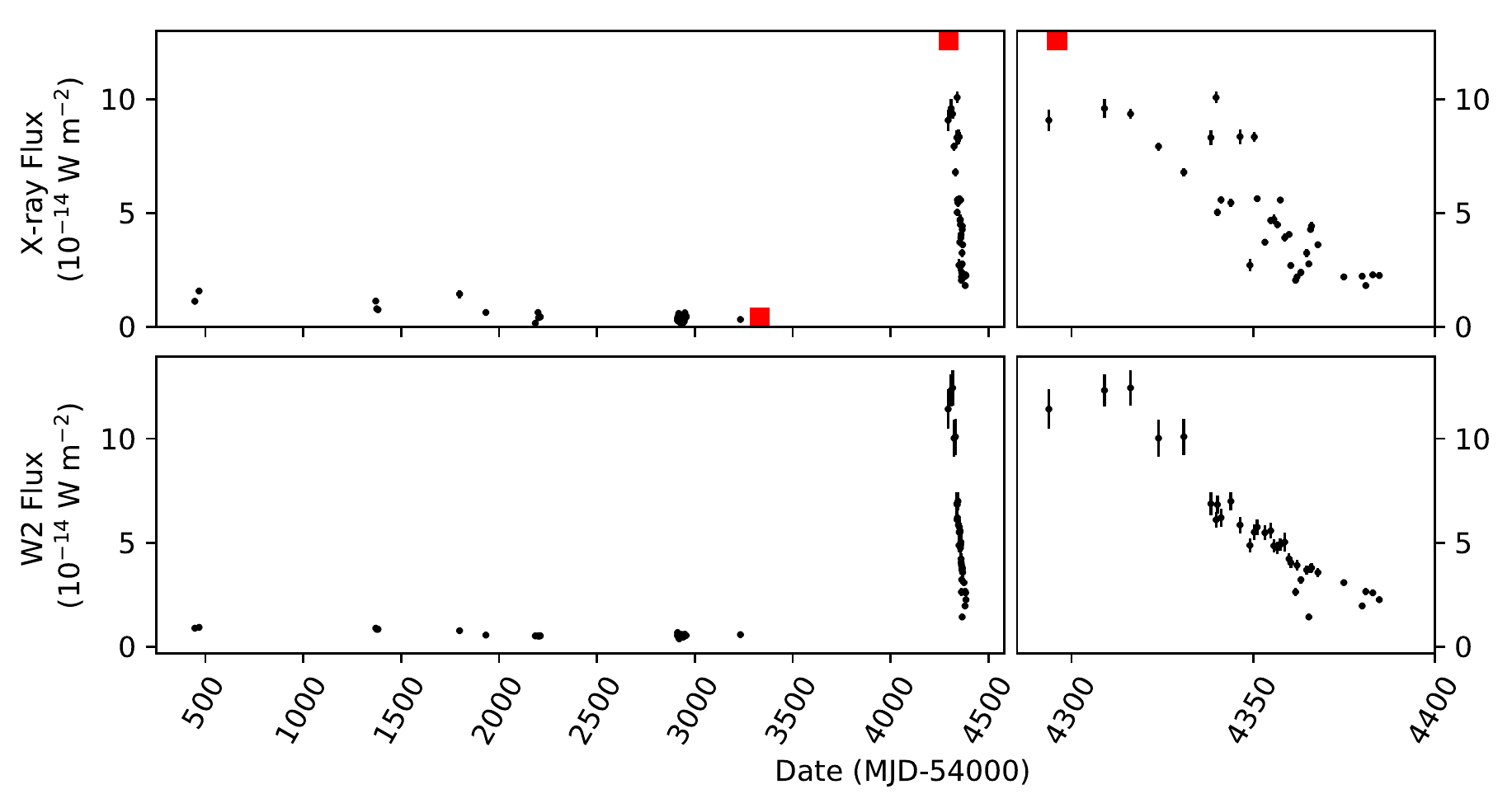}
\caption{Left: Optical monitor (OM) UVW1 image. Right: Long-term \swift\ XRT and UVW2 (2246~\AA) lightcurves. The 2015 and 2018 \xmm\ EPIC-pn fluxes are marked with red squares. Observations started roughly at the peak of the optical outburst and have well sampled the peak and decay. At the time of writing, X-ray flux is around 1/5 of the peak flux, and has been stable for $\sim1$~month.}
\label{fig_lightcurve}
\end{figure*}

\section{Results}

\subsection{RGS}
\label{sec_rgs}
\begin{figure*}
\centering
\includegraphics[height=5.5cm]{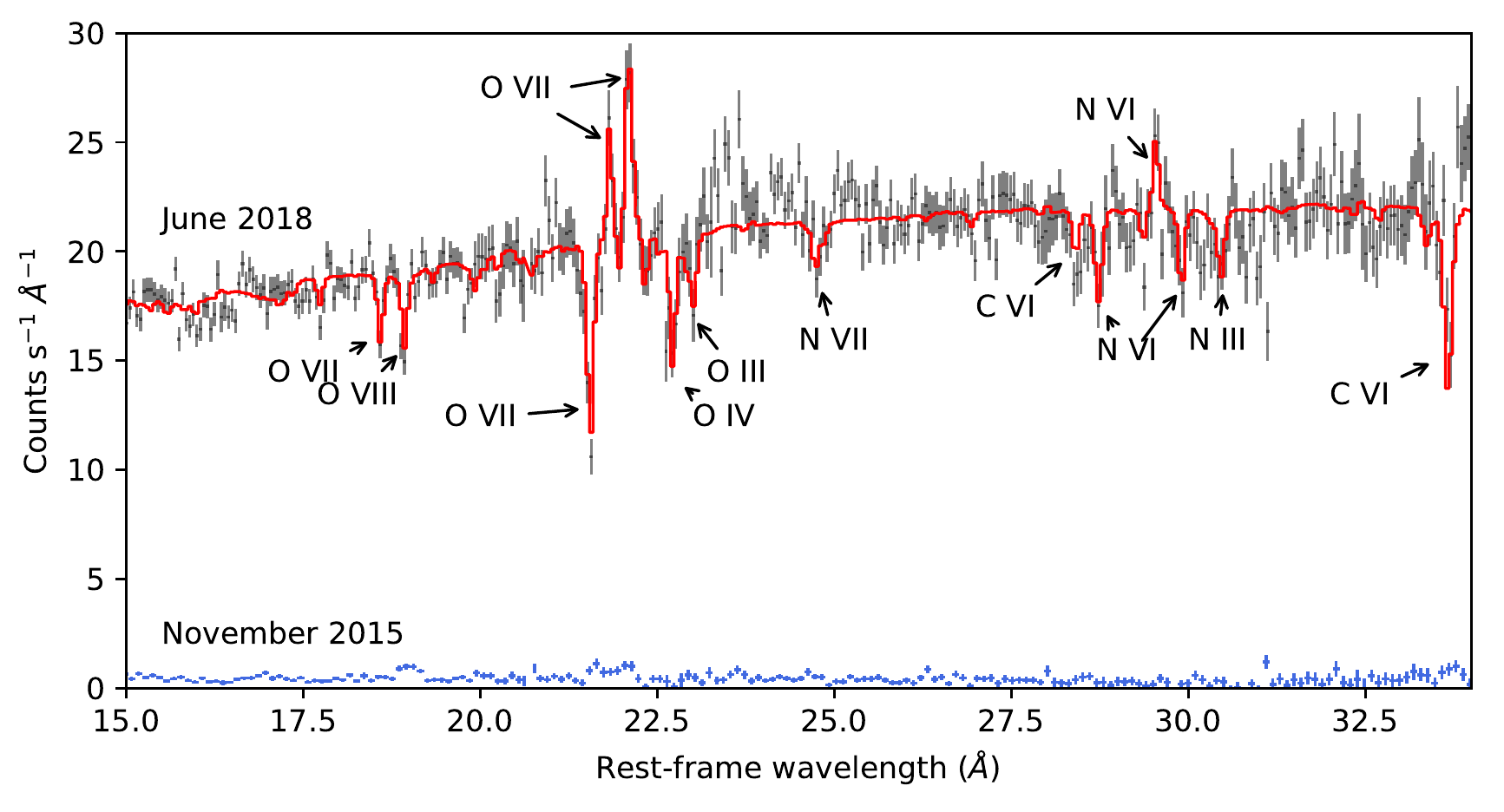}
\includegraphics[height=5.5cm]{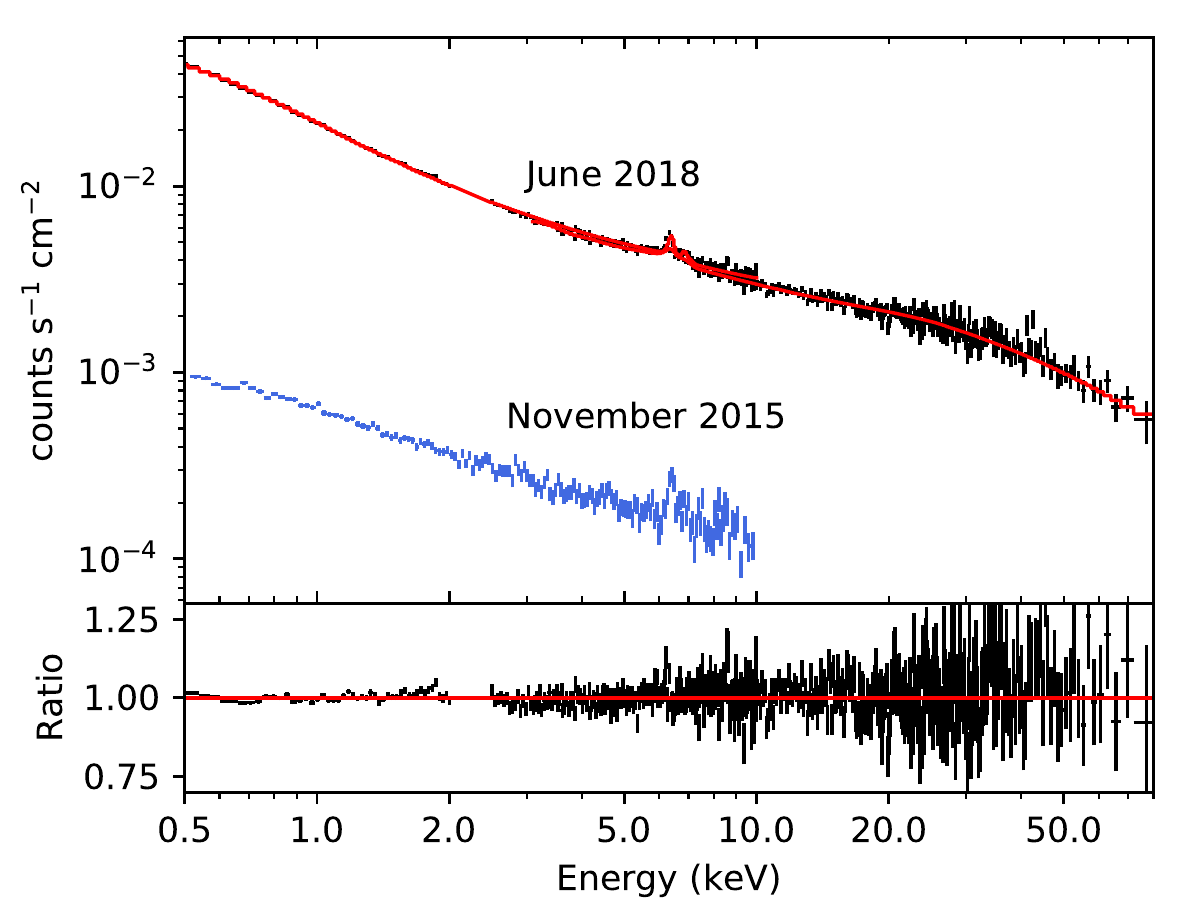}
\caption{Left: RGS spectrum of \src , fit with two zones of warm absorption, and three emission lines. The lower set of points is the corresponding spectrum from 2015, when the source was quiescent. Right: Broad-band \xmm\ and \nustar\ spectrum, corrected for the effective area of the instrument (but not unfolded from the instrumental resolution). The 2018 data is fit with our best-fit model (given in Table~\ref{tab_broad_bestfit}), and the data/model ratio is shown in the lower panel.}
\label{fig_rgs}
\end{figure*}

We initially focus on the RGS spectrum to establish the extent of any absorption in the soft band. We fit the RGS data in \textsc{Spex} \citep{Kaastra96} version 3.03.00. The spectrum shows several clear absorption lines from O, N and C, with an outflow velocity of $\sim500$~\kms , and emission lines at rest from O\textsc{vii} and N\textsc{vi} (Fig.~\ref{fig_rgs}). We fit the spectrum with a simple phenomenological powerlaw plus black-body continuum, two zones of absorption modeled with \emph{xabs}, and three Gaussian emission lines. We also include a \emph{hot} component to model Galactic absorption. This model gives a reasonable description of the data ($\chi^2$/dof $=716/522$), although it misses some structure around 23~\AA. As there are no strong emission lines in this region, this is likely either associated with the O edge, the shape of which can be modified by non-solar abundances or the presence of dust, or due to our simple model not perfectly modeling the continuum.

\begin{table}
\begin{center}
\caption{Best fit parameters for the model fit to the RGS data, shown in Fig.~\ref{fig_rgs}. }
\begin{tabular}{l l c r}
\hline
Comp. & Par. & Value & Description\\
\hline
\emph{hot} 	& $N_\mathrm{H}$ & $(1.2\pm0.3)\times10^{20}$~\pcm & Galactic column\\
			& $kT$	&	$0.25\pm0.03$~keV & Temperature\\
\emph{bb}	& $kT$	&	$0.099\pm0.002$~keV & Temperature\\	
\emph{pow}	& $\Gamma$	&	$2.14_{-0.03}^{+0.02}$	& Photon index\\
\emph{xabs$_1$}	&	$N_\mathrm{H}$	&	$(2.5\pm0.4)\times10^{20}$~\pcm	&	Column\\
				&	$\xi$		&	$10^{-0.7\pm0.1}$ erg~cm~s$^{-1}$&	Ionisation\\
                &	$v_\mathrm{RMS}$	&	$65_{-18}^{+32}$~km~s$^{-1}$	&	RMS velocity\\
                &	$v$		&	$541_{-85}^{+96}$~km~s$^{-1}$	&	Outflow velocity\\
\emph{xabs$_2$}	&	$N_\mathrm{H}$	&	$(2.3\pm0.3)\times10^{20}$&	Column\\
				&	$\xi$		&	$10^{1.2\pm0.1}$ erg~cm~s$^{-1}$&	Ionisation\\
                &	$v_\mathrm{RMS}$	&	$185_{-28}^{+35}$~km~s$^{-1}$&	RMS velocity\\
                &	$v$				&	$472_{-39}^{+43}$~km~s$^{-1}$&	Outflow velocity\\
\emph{gauss$_1^*$}&	$\lambda$	&	$21.83\pm0.01$~\AA	& Wavelength\\
				&	norm	&	$8\pm1$~s$^{-1}$	& Photon flux\\
\emph{gauss$_2^*$}&	$\lambda$	&	$22.19\pm0.01$~\AA	& Wavelength\\
				&	norm	&	$18\pm2$~s$^{-1}$	& Photon flux\\
\emph{gauss$_3^*$}&	$\lambda$	&	$29.69\pm0.02$~\AA	& Wavelength\\
				&	norm	&	$5\pm2$~s$^{-1}$	& Photon flux\\
\hline
\end{tabular}
\end{center}
$^*$The three Gaussian lines correspond to the O\textsc{vii} intercombination and forbidden lines, and the N\textsc{vi} forbidden line, respectively.
\end{table}

The entire RGS spectrum, including the emission lines, is far above the quiescent 2015 spectrum, shown in the lower set of points, so must be dominated by the AGN with negligible contribution from extended emission.


\subsection{Broad-band}
We now fit the broad-band 0.5--79~keV EPIC-pn and \nustar\ spectrum. We fit these data in \textsc{xspec} \citep{Arnaud96} version 12.9.1p. We include Galactic absorption using \emph{tbnew} \citep{Wilms00}, with the column fixed at the value of $9\times10^{19}$~\pcm\ \citep{Kalberla05} consistent with the RGS spectrum.
To account for the warm absorption identified in section~\ref{sec_rgs}, which cannot be directly constrained using the EPIC-pn spectrum, we write the \textsc{spex} model to a text file and convert it to an \textsc{xspec} table model (with no free parameters) using the \emph{flx2tab} \textsc{ftool}. We also add a narrow Gaussian line at 0.566~keV to account for the O\textsc{vii} emission lines, which are unresolved. 


Preliminary fitting from 3--10~keV with a power-law plus distant reflection \citep[modelled with \emph{xillver,}][]{Garcia13} leaves some residuals around the Fe line, which are likely due to relativistic reflection from the accretion disk. We therefore fit the broad-band spectrum with the \emph{relxill} relativistic reflection model \citep{Garcia14}. As this component is weak, we fix the emissivity index to the classical value of 3. We tie the parameters of \emph{xillver} to those of \emph{relxill}, and include an additional soft excess component modelled with \emph{nthcomp} \citep{Zdziarski96,Zycki99}. Finally, we add a Gaussian line at $\sim6.9$~keV to account for a narrow residual. We allow the photon index to vary between the three instruments, and include a constant multiplicative offset between them (the difference between \xmm\ and \nustar\ is large, due to the annular extraction region for the pn).
This model gives a good overall fit ($\chi^2$/dof=818/785), with no strong residuals. This is shown in Fig.~\ref{fig_rgs}, and the parameters are given in Table~\ref{tab_broad_bestfit}.

\begin{table}
\centering
\caption{Best-fit parameters for the broad-band model shown in Fig.~\ref{fig_rgs}.}
\label{tab_broad_bestfit}
\begin{tabular}{l l c r}
\hline
Comp. & Par. & Value & Description\\
\hline
\emph{nthcomp}	&	$kT$	&	$0.8\pm0.1$~keV	&	Temperature\\
				&	$\Gamma$&	$2.69\pm0.02$	&	Photon Index\\	
                & 	norm	&	$0.595\pm0.001$	&	Normalization\\
\emph{relxill}	&	$a$	&	$<0.25$	&	Spin\\
				&	$i$	&	$<11$\textdegree	&	Inclination\\
                &	$\xi$	&	$10^{2.4\pm0.1}$ erg~cm~s$^{-1}$	&	Ionization\\
                &	$A_\mathrm{Fe}$	&	$3.0\pm0.2$	&	Iron abundance\\
                &	$R$		&	$0.091_{-0.004}^{+0.005}$	&	Reflection fraction\\
				&	$\Gamma_\mathrm{pn}$	&	$1.435\pm0.003$&	Photon index\\
				&	$\Gamma_\mathrm{FPMA}$	&	$1.624\pm0.004$&	Photon index\\
				&	$\Gamma_\mathrm{FPMB}$	&	$1.599\pm0.004$&	Photon index\\
                &	$E_\mathrm{cut}$	&	$167\pm3$~keV	&	Cutoff energy\\
                &	norm	& 	($3.89\pm0.01)\times10^{-4}$	&	Normalization\\
\emph{xillver}	&	norm	&	($7.6_{-0.3}^{+0.4})\times10^{-5}$	&	Normalization\\
\emph{zgauss}	&	$E$	&	$6.85^{+0.04}_{-0.05}$~keV	& Energy\\
				&	norm	&	$(1.6\pm0.3)\times10^{-5}$	&	Normalization\\
\emph{const}	&	$C_\mathrm{FPMA}$		&	$0.820\pm0.003$&	Constant offset\\
				&	$C_\mathrm{FPMB}$		&	$0.844\pm0.003$&	Constant offset\\
\hline
\end{tabular}
\end{table}

We estimate Eddington ratios for the 2015 and 2018 X-ray spectra using the 2--10~keV fluxes and a mass of $\sim10^{7}$ \citep{Woo02} and assuming a bolometric correction factor of 20 \citet{Vasudevan09}. This gives Eddington ratios of $\sim$0.2\% for 2015 and $\sim$5\% for 2018.

\section{Discussion}
Overall, the X-ray spectrum of \src\ is not unusual for a Sy~1 galaxy, showing standard spectral components. The ionization and velocity of the outflow are well within the normal range seen in other AGN \citep{Laha14}. The column density is lower than generally seen, but this is likely due to detection bias: in more distant sources, a low column density outflow is unlikely to be detected.
The rapid appearance of this spectrum  after a period of quiescence is very unusual, and the increase in brightness by a factor of 30--40 within a short time period is extreme.
We do not have X-ray data covering the rise period, but the ASAS-SN V-band lightcurve shows that the source flux began to rise around September 2017, peaking around the time of our observation \citep{NGC1566_atel4}. This means the outburst took $\sim9$~months to reach the peak. This is longer than previous outbursts: \citet{Alloin86} report a typical rise time of $\sim20$~days, and the outburst visible in the 105~month \swift\ BAT catalog reaches the peak in 3~months. Given the lack of X-ray coverage of previous outbursts, we cannot explain this difference from an X-ray perspective without further data.

Changing look events like this one involve flux changes  in multiple wavebands on timescales far faster than a standard thin disk can evolve.
There are several different mechanisms that are invoked to explain this phenomenon, such as obscuration, disk instabilities, and tidal disruption events (TDEs). Variable obscuration, where clumps of cold gas from the torus block the AGN emission \citep{Matt03}, can be discounted in this case as the optical broad lines have responded directly to the increased optical and UV flux. 

Large changes in accretion rate can produce large changes in the flux at all wavelengths, but for a `standard' disk the timescales involved are far too long. \citet{Dexter18} show that disks supported by magnetic pressure have much faster infall times, and can produce changes of a factor of 2--10 in optical to X-ray flux within 1--10 years. While promising for many changing-look events, this is likely not extreme enough to reproduce the rapid increase in flux seen in \src , which brightened by a factor of $\sim40$--70 and has had previous outbursts with rise times of less than a month.

In principle, TDEs can produce repeated events over many years (for example by repeated tidal stripping of a star), and this has been suggested as a possible explanation of repeat X-ray flares in IC~3599 \citep[][]{Campana15}. However, we consider this unlikely in this case. The theoretical rate of TDEs is low ($\sim10^{-4}$ per Galaxy, per year), and to find one in such a nearby galaxy that already hosts an AGN would be very unusual. The similarity of this outburst to other changing look events, which are common in nearby galaxies \citep[e.g.]{Runco16} at a rate far in excess of the predicted TDE rate, suggests a common non-TDE origin. Finally, the X-ray spectrum of \src\ is a classic hard AGN spectrum, whereas X-ray spectra of TDEs are typically extremely soft \citep[][and references therein]{Komossa17_TDEs}.

In our view, the most likely interpretation of this behaviour is an instability in the accretion disk. \citet{Grupe15_ic3599} discuss this for IC~3599, a Sy~1.9 AGN that has undergone at least two large outbursts. \citet{Saxton15} and \citet{Grupe15_ic3599} explore the \citet{Lightman74} instability, where the inner disk is quiescent until radiation pressure exceeds gas pressure, at which point the disk rapidly switches on. This mechanism produces variability on around the right timescales but requires that the rise time be longer than the decay time, a condition which has not been met by previous outbursts in \src. Additionally, the repeat time is set by the viscous time at the truncation radius which is typically decades, much longer than observed in \src\ \citep{Alloin86}.
\citet{Ross18} explain the changing look of a $z\sim0.4$, $M\sim10^{8.8}M_\odot$ quasar with a cooling front that propagates away from the ISCO, followed by a returning heating front over 20 years. This is similar in timescale after scaling for the mass ($\sim$ a few months), but predominantly affects the flux at short wavelengths, so does not explain the  uniform flux increase and decline in \src .
\citet{Noda18} propose that the drop in flux by a factor of 10 in Mrk~1018 and associated change from Sy~1.9 to Sy~1 is caused by a combination of the H~instability, which produces the overall drop in luminosity, and evaporation of the inner disk, which causes the associated spectral hardening. These processes work in reverse, so a heating front caused by the H~instability could propagate through the disk, causing an outburst. Interestingly, \citeauthor{Noda18} suggest that sources crossing a few per cent of Eddington should go through a changing look along with strong soft excess variability, and no soft excess was observed in earlier quiescent observations of \src\ \citep{Kawamuro13}. The timescales of the changing look events in these two sources are similar when the higher mass of Mrk~1048 \citep[$M\sim10^{7.8}M_\odot$, see][]{Noda18} is taken into account (9 months $\times10\sim8$ years), although the variability amplitude is greater in \src . Given these similarities between the outbursts, we consider it likely that they are due to the same mechanism.

The inclination measured from the relativistic Fe line is very low ($<11$\textdegree), which is consistent with the accretion disk being aligned with the face-on galaxy. While this is expected, we note that \citet{Middleton16_inclinations} found that there is not a strong correlation between host and disk inclinations. We find only an upper limit on the spin, of $<0.25$. A poor constraint is to be expected, given the weak feature and low inclination, which gives a narrow, hard to measure line. The low spin is interesting, and may be indicating some truncation of the accretion disk \citep[although there are other explanations, e.g.][]{Parker18_tons180}. While these results are intriguing, it is not possible to come to robust conclusions because of the limited signal available.


\section{Conclusions}
In 2018, the nearby Seyfert galaxy \src\ went through a major outburst, reaching a peak X-ray flux in excess of 70 times the minimum observed with the \swift\ XRT, before slowly decaying by a factor of $\sim5$ at the time of writing. This coincided with a change in Seyfert type to 1.2. We triggered a joint \xmm\/\nustar\ observation at the peak of the outburst, to obtain an X-ray perspective, and have uncovered several key results.
\begin{itemize}
\item The high resolution RGS spectrum shows several absorption lines from ionized gas, outflowing at $\sim500$~\kms . This outflow was likely launched or accelerated by the large increase in radiation pressure from the AGN outburst.
\item There are several emission lines from ionized O, N, and Fe at rest in the \xmm\ data, produced by photoionization of cold gas somewhere in the AGN system, such as the outer disk, torus, or BLR clouds. These lines are strong, which may indicate a significant solid-angle of cold gas.
\item The broad-band \xmm/\nustar\ spectrum requires a weak contribution from relativistic reflection off the accretion disk. The inclination is low ($<11$\textdegree), consistent with the face-on inclination of the galaxy.
\end{itemize}
Overall, the X-ray spectrum is not unusual for a Sy~1 AGN, however this is itself noteworthy given the extreme nature of the outburst. This implies that a `standard' accretion disk can develop very quickly. We discuss possible mechanisms for the outburst, and conclude that the most likely scenario is a disk instability that rapidly and uniformly increases the emission from the AGN. \src\ is an excellent candidate for understanding the changing-look phenomenon, due to its proximity and repeat outburst behaviour.

\section*{Acknowledgements}
MLP is supported by an ESA Research Fellowship. We thank the anonymous referee for detailed and constructive feedback. We thank J. Lewis for editing the manuscript. Based on observations obtained with \textit{XMM-Newton}, an ESA science mission funded by ESA Member States and NASA, and the \textit{NuSTAR} mission, a project led by Caltech, managed by JPL, and funded by NASA. This work has been supported by the DFG grant Ko 857/33-1. We thank the \textit{Swift} team for carrying out our ToO requests.



\bibliographystyle{mnras}
\bibliography{bibliography_ngc1566} 





\bsp	
\label{lastpage}
\end{document}